\documentclass[aps,reprint,groupaddress]{revtex4-1}
\usepackage{graphicx}
\usepackage{braket}
\usepackage{bm}
\usepackage{amsmath}
\begin{document}
\title{3$\bm \alpha$ clustering in the excited states of $^{\bm{16}}$C}
\author{T. Baba, Y. Chiba and M. Kimura}
\affiliation{Department of Physics, Hokkaido University, 060-0810 Sapporo, Japan}
\date{\today}

\begin{abstract}
 The $\alpha$ cluster states of $^{16}$C are investigated by using the  antisymmetrized  molecular
 dynamics. It is shown that two different types of $\alpha$ cluster states exist:  triangular and
 linear-chain states.  The former has an  approximate isosceles triangular  configuration of  $\alpha$
 particles surrounded by  four valence  neutrons  occupying $sd$-shell, while the latter has the 
 linearly  aligned  $\alpha$  particles with  $(sd)^2(pf)^2$ neutrons. It is found that the structure of 
 the linear-chain state is qualitatively understood in terms of the   $3/2^-_\pi$ and  $1/2_\sigma^-$
 molecular-orbit  as  predicted  by  molecular-orbital model, but there exists non-negligible
 $^{10}$Be+$\alpha$+2$n$  correlation. The band-head energies  of the triangular and  linear-chain
 rotational bands  are 8.0 and 15.5 MeV, and the latter is close to the  $^{4}$He+$^{12}$Be and
 $^{6}$He+$^{10}$Be  threshold  energies. It is also shown that the  linear-chain state becomes  the  yrast
 state at  $J^\pi=10^+$ with  $E_x=27.8$ MeV  owing to its very large moment-of-inertia comparable with
 hyperdeformation.   
\end{abstract}

\maketitle
\section{introduction}
The excited states of atomic nuclei, especially those of light nuclei, show strong $\alpha$ clustering, and
many different types of $\alpha$ cluster structure appear as the excitation energy increases
\cite{wild58,shel60,abe80,ortz06,hori12}. In particular, the  linear-chain configuration of three $\alpha$
particles (linearly aligned three $\alpha$ particles) suggested by Morinaga \cite{mori56} has long been an
important and interesting subject because of its exotic structure and large deformation equivalent to
hyperdeformation. The Hoyle state ($0^+_2$ state of $^{12}$C) was the first candidate of the linear chain,
but later it turned out that it does not have the linear-chain configuration but are loosely coupled
3$\alpha$ particles with dilute gas-like nature \cite{uega77,kami81,tohs01,funa03}. In turn, the instability
of the linear-chain configuration against the bending motion (deviation from linear alignment) was pointed
out and the bent-armed configuration was predicted by the antisymmetrized molecular dynamics (AMD)
\cite{enyo97} and Fermionic molecular dynamics (FMD) calculations \cite{neff07}. 

The interest in the linear-chain state is reinforced by  the unstable nuclear physics, because
the addition of the valence neutrons will increase the stability of $\alpha$ cluster structure by their
glue-like role. For example, 2$\alpha$ cluster structures of Be isotopes are assisted by valence
neutrons that are well described in terms of the molecular-orbits
\cite{seya81,oert96,itag00,yeny99,amd1,amd2}. Naturally, we expect that the linear-chain configurations of
3$\alpha$ clusters can be stabilized by the assist of valence neutrons in neutron-rich C isotopes. Indeed,
there are a number of studies to theoretically predict and experimentally search for the linear-chain states
in neutron-rich carbon isotopes \cite{itag01,gree02,bohl03,oert04,ashw04,itag06,pric07,suha10,furu11}. 
Among C isotopes, $^{16}$C is very interesting and important nucleus as the most promising candidate of
the stable linear-chain state, because its stability against the bending motion was pointed out by
molecular-orbital model calculation \cite{itag01}. Assuming  3$\alpha$ cluster core and  $3/2^-_\pi$,
$1/2^-_\pi$ and $1/2^-_\sigma$ molecular-orbits of valence neutrons, it was shown  that the linear-chain
configuration with valence neutrons occuping $(3/2^-_\pi)^2(1/2^-_\sigma)^2$ molecular-orbits is
stable. Therefore, it is very important and interesting to investigate the linear-chain state in  $^{16}$C
without a-priori  assumption on the cluster core and valence neutron orbits. Furthermore, in addition to
the linear-chain configuration, triangular configurations of 3$\alpha$ particles are also suggested in the
neighbouring nuclei such as $^{13}$C and $^{14}$C \cite{suha10,furu11,oert03,itag04}. Therefore, it is also
interesting to search for analogous state in $^{16}$C.

For this purpose, we discuss 3$\alpha$ cluster states in $^{16}$C based on AMD which has
been successfully applied to the studies of the clustering in unstable nuclei
\cite{amd1,amd2,furu08,kimu07,kimu11}. Our aim in the present study is two-fold. The first is to search for
and show the linear-chain  and other cluster states in $^{16}$C without a-priori assumption on the structure
and to test the stability against the bending motion. We will show that two different types of the
3$\alpha$  cluster states exist, triangular and linear-chain configurations. It is also shown that the
valence neutron orbits are qualitatively understood in terms of the molecular-orbits, and the linear-chain
configuration is stable with the help of those valence neutrons. The second aim is to provide a quantitative
and reliable prediction of their properties for the experimental survey.  We predict the band-head states of
the triangular and linear-chain bands at 8.0 and 15.5 MeV, and the $J^\pi=10^+$ state of the linear-chain
configuration becomes the yrast state at $J^\pi=10^+$ with  $E_x=27.8$ MeV owing to its very large moment of
inertia comparable with hyperdeformation.

\section{theoretical framework}

\subsection{variational calculation and generator coordinate method}
The microscopic $A$-body Hamiltonian used in this study is written as,
\begin{align}
 \hat{H} = \sum_{i=1}^A \hat{t}(i) + \sum_{i<j}^A \hat{v}_n(ij) + \sum_{i<j}^Z \hat{v}_C(ij)
 - \hat{t}_{c.m.},
\end{align}
where the Gogny D1S interaction \cite{gogn91} is used as an effective nucleon-nucleon interaction
$\hat{v}_n$ and the Coulomb interaction $\hat{v}_C$ is approximated by a sum of seven Gaussians. The
center-of-mass kinetic energy $\hat{t}_{c.m.}$ is exactly removed. 

The intrinsic wave function $\Phi_{int}$ of the system  is represented by a Slater determinant
of single particle wave packets, and we employ the parity-projected wave function  $\Phi^\pi$ as the
variational wave function,
\begin{align}
 \Phi^\pi &= \frac{1+\pi \hat{P}_x}{2}\Phi_{int},\quad
 \Phi_{int} ={\mathcal A} \{\varphi_1,\varphi_2,...,\varphi_A \},
  \label{EQ_INTRINSIC_WF}  
\end{align}
where $\varphi_i$ is the single particle wave packet which is a direct product of the deformed Gaussian
spatial part \cite{kimu04}, spin ($\chi_i$) and isospin ($\xi_i$) parts, 
\begin{align}
 \varphi_i({\bf r}) &= \exp\biggl\{-\sum_{\sigma=x,y,z}\nu_\sigma
  \Bigl(r_\sigma - 
  \frac{Z_{i\sigma}}{\sqrt{\nu_\sigma}}\Bigr)^2\biggr\}\chi_i\xi_i, \label{eq:singlewf}\\
 \chi_i &= a_i\chi_\uparrow + b_i\chi_\downarrow,\quad
 \xi_i = {\rm proton} \quad {\rm or} \quad {\rm neutron}.\nonumber
\end{align}
In this study, we focus on the positive-parity states of $^{16}{\rm C}$.
The parameters ${\bm Z}_i$, $a_i$, $b_i$ and $\nu_\sigma$ are optimized by the variational calculation
explained below. To investigate 3$\alpha$ cluster states, we first perform the variational calculation with
the constraint on the quadrupole deformation parameter $\beta$. In this calculation, we do not impose
constraint on the parameter $\gamma$, and hence, thus-obtained wave functions have $\gamma$ values that give
the largest binding energies for given values of $\beta$. As shown in the next section, we have obtained the
linear-chain configuration located at $(\beta,\gamma)=(1.10,0)$ as well as the triangular
configuration. We performed another variational calculation to test its stability against bending
motion (deviation from the linear alignment of 3$\alpha$ clusters). Namely, starting from the above
mentioned linear-chain configuration, we gradually increased the parameter $\gamma$ keeping $\beta=1.10$ by
applying the constraints on $\beta$ and $\gamma$ simultaneously. This calculation generates the energy curve
of the linear-chain configuration as function of $\gamma$.

After the variational calculation, the eigenstate of the total angular momentum $J$ is projected out from
the wave functions $\Phi^+_i$ obtained by variational calculations,  
\begin{eqnarray}
 \Phi^{J+}_{MKi} = \frac{2J+1}{8\pi^2}\int d\Omega D^{J*}_{MK}(\Omega)\hat{R}(\Omega)\Phi^{+}_i.
\end{eqnarray} 
Here, $D^{J}_{MK}(\Omega)$ is the Wigner $D$ function and $\hat{R}(\Omega)$ is the rotation operator. The
integrals over  three Euler angles $\Omega$ are evaluated numerically. Then, we perform the GCM calculation
by employing the quadrupole deformation parameter $\beta$ as the generator coordinate.  The wave functions
$\Phi^{J+}_{MKi}$ are superposed, 

\begin{align}
 \Psi^{J+}_{M\alpha} = \sum_{Ki}g^{J}_{Ki\alpha}\Phi^{J+}_{MKi},\label{eq:gcmwf}
\end{align}

where the coefficients $g^{J}_{Ki\alpha}$ and eigenenergies $E^{J+}_\alpha$ are obtained by solving the
Hill-Wheeler equation \cite{hill54}, 
\begin{align}
 \sum_{i'K'}{H^{J+}_{KiK'i'}g^{J}_{K'i'\alpha}} &= E^{J+}_\alpha \sum_{i'K'}{N^{J+}_{KiK'i'}g^{J}_{K'i'\alpha}},\\
  H^{J+}_{KiK'i'} &= \braket{\Phi^{J+}_{MKi}|\hat{H}|\Phi^{J+}_{MK'i'}}, \\
  N^{J+}_{KiK'i'} &= \braket{\Phi^{J+}_{MKi}|\Phi^{J+}_{MK'i'}}.
\end{align}
The wave functions $\Psi^{J+}_{M\alpha}$ that describe the ground and excited states of $^{16}$C are called 
GCM wave function in the following.

\subsection{single particle orbits}
To investigate the motion of the valence neutrons around the core nucleus, we calculate the neutron
single-particle orbits of the intrinsic wave function. We first transform the single particle wave packet
$\varphi_i$ of each optimized intrinsic wave function $\Phi_{int}$ to the orthonormalized basis,
\begin{align}
 \widetilde{\varphi}_\alpha = \frac{1}{\sqrt{\lambda_\alpha}}\sum_{i=1}^{A}c_{i\alpha}\varphi_i.  
\end{align}
Here, $\lambda_\alpha$ and $c_{i\alpha}$ are the eigenvalues and eigenvectors of the
overlap matrix $B_{ij}=\langle\varphi_i|\varphi_j\rangle$. Using this basis, the
 Hartree-Fock single particle Hamiltonian is derived,
\begin{align}
 h_{\alpha\beta} &=
  \langle\widetilde{\varphi}_\alpha|\hat{t}|\widetilde{\varphi}_b\rangle + 
  \sum_{\gamma=1}^{A}\langle
  \widetilde{\varphi}_\alpha\widetilde{\varphi}_\gamma|
  {\hat{v}_n+\hat{v}_C}|
  \widetilde{\varphi}_\beta
\widetilde{\varphi}_\gamma -
\widetilde{\varphi}_\gamma\widetilde{\varphi}_\beta\rangle,\nonumber\\ 
 &+\frac{1}{2}\sum_{\gamma,\delta=1}^{A}
 \langle\widetilde{\varphi}_\gamma\widetilde{\varphi}_\delta 
|\widetilde{\varphi}_\alpha^*\widetilde{\varphi}_\beta
\frac{\delta\hat{v}_n}{\delta \rho}|\widetilde{\varphi}_\gamma
\widetilde{\varphi}_\delta - \widetilde{\varphi}_\delta  \widetilde{\varphi}_\gamma
\rangle.
\end{align}
The eigenvalues $\epsilon_s$ and eigenvectors  $f_{\alpha s}$ of $h_{\alpha\beta}$ give the single particle
energies and the single particle orbits,
 $\widetilde{\phi}_s = \sum_{\alpha=1}^{A}f_{\alpha s}\widetilde{\varphi}_\alpha$.  To discuss the properties
 of the single particle levels, we also calculate the amount of the positive-parity component, 
\begin{align}
 p^+ = |\langle \widetilde{\phi}_s|\frac{1+P_x}{2}| \widetilde{\phi}_s\rangle|^2, \label{eq:sp1}
\end{align}
and angular momenta in the intrinsic frame,
\begin{align}
 j(j+1)&= \langle \widetilde{\phi}_s|\hat{j}^2| \widetilde{\phi}_s\rangle, \quad
 |j_z| = \sqrt{\langle \widetilde{\phi}_s|\hat{j}_z^2| \widetilde{\phi}_s\rangle},\label{eq:sp2}\\
 l(l+1)&= \langle \widetilde{\phi}_s|\hat{l}^2| \widetilde{\phi}_s\rangle, \quad
 |l_z| = \sqrt{\langle \widetilde{\phi}_s|\hat{l}_z^2| \widetilde{\phi}_s\rangle}.\label{eq:sp3}
\end{align}

\section{results and discussions}
\subsection{3$\bm \alpha$ clustering and valence neutron configurations on the energy curve} 
\begin{figure}[h]
 \includegraphics[width=1.0\hsize]{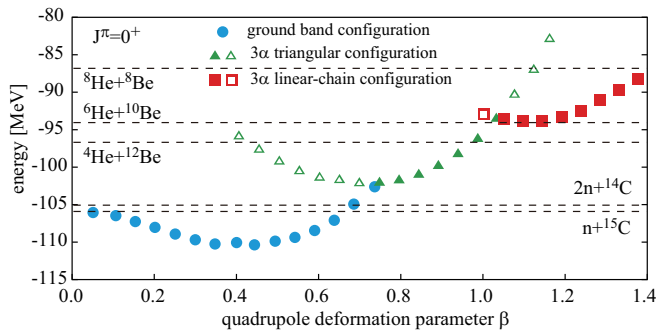}
 \caption{(color online) The energy curve of the $J^\pi=0^+$ states as functions of quadrupole deformation
 parameter $\beta$ obtained  by the angular momentum projection. Filled symbols show the energy
 minimum states for given values of $\beta$, while open symbols show local energy minima. There appears
 three different structures shown by circles, triangles and boxes (see text). Dashed lines show the
 thresholds energies for 1$n$, $2n$ and cluster decays.} 
 \label{fig:surface}
\end{figure}
\begin{figure}[h]
 \centering
 \includegraphics[width=1.0\hsize]{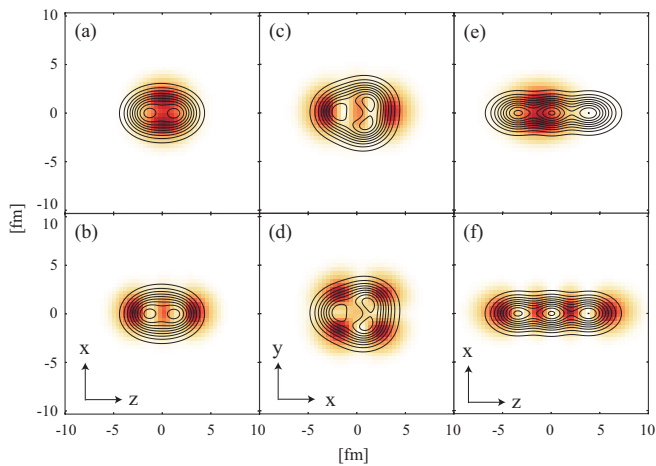}
 \caption{(color online) The density distribution of the ground (a)(b), triangular (c)(d) and linear-chain
 (e)(f)  configurations at their energy minima. The contour lines show the proton density distributions and
 are  common to the upper and lower panels. The colour plots show the single particle orbits occupied by
 four  valence neutrons. The lower panels show the most weakly bound two neutrons, while the upper panel
 show the  other two valence neutrons.} 
 \label{fig:density}
\end{figure}
\begin{figure}[h]
 \centering
 \includegraphics[width=1.0\hsize]{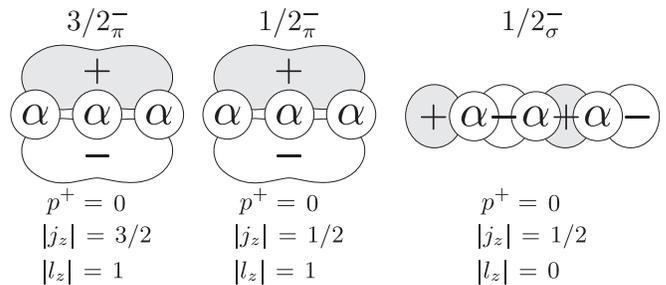}
 \caption{The schematic figure showing the $3/2^-_\pi, 1/2^-_\pi$ and $1/2^-_\sigma$ molecular orbits
 introduced in Ref. \cite{itag01}. If the system has axial symmetry and the effect of the spin-orbit
 interaction is negligible, these orbits are the eigenstates of $\hat j_z$ and $\hat l_z$.} 
 \label{fig:illust}
\end{figure}

\begin{table}[h]
\caption{The properties of valence neutron orbits shown in Fig. \ref{fig:density}. Each column show the
 single particle energy $\varepsilon$ in MeV, the amount of the positive-parity component $p^+$ and the
 angular momenta (see Eqs. (\ref{eq:sp1})-(\ref{eq:sp3})).}
\label{table:spo}
\begin{center}
 \begin{ruledtabular}
  \begin{tabular}{ccccccc} 
	orbit & $\varepsilon $ & $p^+$ & $j$ & $|j_{z}|$ & $l$ & $|l_{z}|$ \\ \hline
	(a) & $-8.24$ & 0.00 & 0.75 & 0.51 & 1.05 & 0.97 \\
	(b) & $-5.23$ & 0.99 & 2.21 & 0.51 & 1.80 & 0.38 \\ \hline
	(c) & $-5.74$ & 0.99 & 2.31 & 1.96 & 1.93 & 1.63 \\
	(d) & $-3.29$ & 0.98 & 2.33 & 1.88 & 2.07 & 1.83 \\ \hline
	(e) & $-5.32$ & 0.13 & 2.09 & 1.49 & 1.72 & 0.99 \\
	(f) & $-4.18$ & 0.03 & 2.89 & 0.53 & 2.72 & 0.18 \\ 
  \end{tabular}
  \end{ruledtabular}
\end{center}
\end{table}

Figure \ref{fig:surface} shows the energy curves as functions of quadrupole deformation parameter $\beta$
for $J^\pi=0^+$ states obtained by the variational calculation with the constraint on the parameter
$\beta$. The filled symbols show the energy minimum for given values of $\beta$, and on this energy curve,
three different structures appear which are shown by circles, triangles and boxes. These structures are also
obtained as the local energy minima above the lowest energy states, and are shown by open symbols. It is
also noted that there are other local energy minima with different structure above the energy curve. They do
not have cluster structure and are not shown in Fig. \ref{fig:surface}, but included as the basis wave
function of the GCM calculation. We first discuss three different structures with and without clustering
that appear on the energy curve by referring their density distributions (Fig. \ref{fig:density}) and the
properties of valence neutron  orbits (Tab. \ref{table:spo}). The lowest energy configuration shown by
circles is prolately deformed and has the minimum at $E=-110.4$ MeV and $(\beta,\gamma)=(0.44,0)$. As seen
in its proton and valence neutron density distribution (Fig.\ref{fig:density} (a) and (b)), it has no
pronounced clustering, and four valence neutrons have an approximate $(0p_{1/2})^2(0d_{5/2})^2$
configuration that is also confirmed from the properties of neutron single particle orbits listed in
Tab. \ref{table:spo} (a) and (b). Namely the first two valence neutrons occupy the orbit (a) with negative
parity, $j\simeq 1/2$ and $l\simeq 1$, and the last two neutrons occupy the orbit (b) with positive parity,
$j\simeq5/2$ and $l\simeq2$. The deviation from the spherical $p_{1/2}$ and $d_{5/2}$ orbits owes to 
prolate deformation of this state. Different from the AMD results by Kanada-En'yo \cite{yeny05}  
in which the different proton and neutron deformation of $^{16}$C was discussed ({\it i.e.} proton is
oblately deformed, while neutron is prolately deformed), the present result shows that the both proton
and neutron are prolately deformed in the ground state. This difference may be attributed to the difference
of the basis wave functions used in this study and Ref. \cite{yeny05}. In the present study, we use the
deformed Gaussian (Eq. \ref{eq:singlewf}) whose deformation is {\it common} to protons and neutrons, as a
result, the different deformation between proton and neutron may be energetically unfavoured.

As deformation increases, other valence neutron configuration appears and it induces 3$\alpha$
clustering. A triaxially deformed 3$\alpha$ cluster configuration shown by triangles appears around
$\beta=0.7$ and has the local energy minimum at $E=-102.2$ MeV and $(\beta,\gamma)=(0.70,41)$. At the energy
minimum, this configuration has 3$\alpha$ cluster core of an approximate isosceles triangular configuration
with 3.2 fm long sides and 2.3 fm short side (Fig. \ref{fig:density} (c) and (d)) which is the origin of the
triaxial deformation, and an approximate $(0d_{5/2})^4$ configuration ($2\hbar\omega$ excitation) of valence
neutrons are confirmed from Tab. \ref{table:spo}.  It is also notable that $|j_z|$ of valence neutron
orbits deviate from half-integer value because of axial symmetry breaking caused by the triangular
configuration. Thus, by increasing the nuclear deformation, valence neutron configuration changes and it
triggers the clustering of the core nucleus. This feature is common to the well-known 2$\alpha$ clustering
of Be isotopes and theoretically predicted clustering in O, F and Ne \cite{furu08,kimu07,kimu11} isotopes.

Further increase of nuclear deformation realizes the exotic cluster configuration with the linear
alignment of 3$\alpha$ particles which is denoted by boxes. This configuration has a local minimum at
$E=-93.9$ MeV and $(\beta,\gamma)=(1.10,0)$ whose energy is very close to the $^{4}$He+$^{12}$Be and
$^{6}$He+$^{10}$Be cluster thresholds, and the ratio of deformation axis is approximately equal to 3:1. As
clearly seen in Fig. \ref{fig:density} (e) and (f), a linearly aligned 3$\alpha$ cluster core is
accompanied by four valence neutrons whose configuration may be roughly understood as $(1p)^2(0f)^2$,
although the deviation from ordinary spherical shell is fairly large due to very strong deformation. 
An alternative and more appropriate interpretation of the valence neutron configuration is given by the
molecular orbits. Namely, the valence neutron orbits are in good accordance with the 
$3/2^-_\pi$ and $1/2^-_\sigma$ orbits \cite{itag01} that are the linear combinations of the
$p$-orbits around $\alpha$ clusters as illustrated in Fig. \ref{fig:illust}. Indeed, the density
distribution and properties of these orbits shown in Fig. \ref{fig:density} are in very good agreement
with those of the molecular-orbital model.  It is also noted that the $(3/2^-_\pi)^2(1/2^-_\pi)^2$
configuration was not obtained in this study, and hence, the present results support the instability of
$(3/2^-_\pi)^2(1/2^-_\pi)^2$ configuration and stability of $(3/2^-_\pi)^2(1/2^-_\sigma)^2$ configuration.  

Thus, concerning the linear-chain configuartion of $^{16}$C,  the present calculation yielded qualitatively
the same conclusion with the molecular-orbital model. However, it is worthwhile to focus on the
quantitative differences. The linear-chain configuration
obtained in this study has parity asymmetric structure and shows $^{10}{\rm Be}+\alpha+2n$ like correlation,
which is analogous to $^{10}{\rm Be}+\alpha$ correlation in $^{14}$C reported by Suhara {\it et al.}
\cite{suha11}. Namely, the $3/2^-_\pi$-orbit has non-negligible parity mixing ($p^+=0.13$) and is localized 
between the left and center $\alpha$ clusters showing  similar  structure to $^{10}$Be. Indeed, owing to
the glue-like role of $3/2^-_\pi$ orbit, the distance between the left and center $\alpha$ clusters (3.5
fm) is shorter than that between the right and center (3.8 fm). On the other hand, the $1/2^-_\sigma$
orbit has almost no parity mixing ($p^+=0.03$) and distributes around the entire system to 
bond  $^{10}{\rm Be}$ and $\alpha$ clusters. Therefore, this state can be alternatively interpreted as
$^{10}{\rm Be}+\alpha$ clusters accompanied by two covalent neutrons in $1/2^-_\sigma$-orbit. This
interpretation may explain why the excitation energy of the linear-chain configuration is is much lower than
that predicted by the molecular-orbital model and located in the vicinity of the $^{6}$He+$^{10}$Be and
$^{4}$He+$^{12}$Be  thresholds. It is evident that the parity-projection plays a crucial role to yield this
asymmetric internal structure, because we only obtain parity-symmetric intrinsic wave functions if we do not
perform parity-projection.

\subsection{stability of the linear-chain state}
One of the main concerns about the linear-chain configuration is its stability against the bending motion,
and we confirm it by investigating  its response to $\gamma$ deformation.  

\begin{figure}[h]
 \includegraphics[width=\hsize]{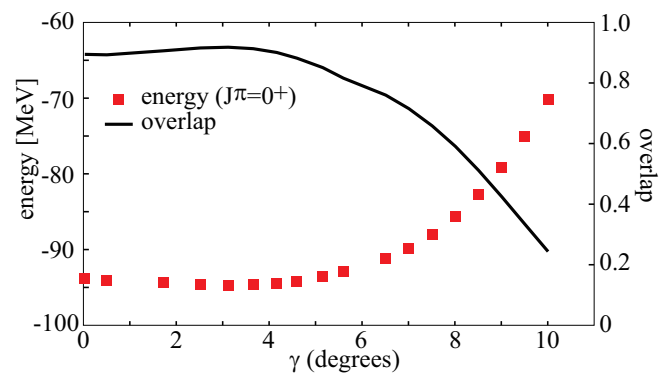}
 \caption{(color online) The boxes show the energy of the linear-chain configuration with  $J^\pi=0^+$
 as function of quadrupole deformation parameter $\gamma$. The solid line shows the overlap
 between the linear-chain state ($0^+_5$ state) and the basis wave functions.} 
 \label{fig:gamma} 
\end{figure}

Starting from the linear-chain configuration shown in Fig. \ref{fig:density} (e)(f), we gradually increased 
$\gamma$ but kept $\beta$ constant by using the constraint on $\beta$ and
$\gamma$. Thus obtained energy curve of the linear-chain configuration with $J^\pi=0^+$ as function of
$\gamma$ is shown in Fig. \ref{fig:gamma}. It is almost constant for small value of $\gamma$ and has the
minimum at $\gamma=3.1$ degrees, but rapidly increases for larger values of $\gamma$. Then, including all
the  basis wave functions, we performed GCM calculation to obtain the excitation spectrum and band structure
which will be discussed in the next subsection. Here, we focus on the band-head state of the linear-chain
band ($0^+_5$ state) to see the stability against $\gamma$ deformation. For this purpose, we calculated the
overlap between the $0^+_5$ state  and the  basis wave function with $\gamma$ deformed linear-chain
configuration defined as, 
\begin{align}
 O(\gamma) = |\langle\Psi(0^+_5)|\Phi^{0^+}(\gamma)\rangle|^2.
\end{align}
Here, $\Psi(0^+_5)$ and $\Phi^{0^+}(\gamma)$ denote the GCM wave function of the $0^+_5$ state and the basis
wave function with $\gamma$ deformed linear-chain configuration shown in Fig. \ref{fig:gamma}.
The calculated overlap shown by the solid line in Fig. \ref{fig:gamma} has its maximum value 0.92 at
$\gamma=3.1$  degrees and falls off very quickly as $\gamma$ increases. Therefore the wave function of the
linear-chain state is well confined within a region of small $\gamma$, and hence stable against the bending
motion. Further extensive investigation of the stability of the linear-chain state including other carbon
isotopes will be discussed in our forthcoming paper.  
\subsection{excitation spectrum}
\begin{figure*}[t]
 \includegraphics[width=\hsize]{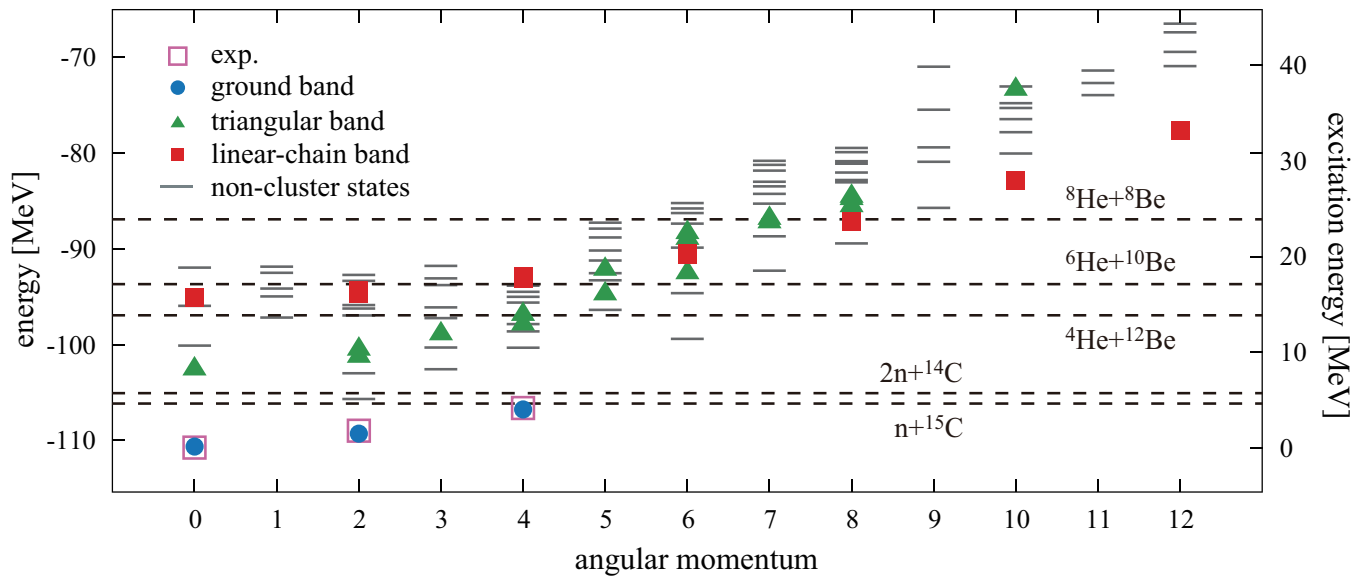}
 \caption{(color online) The calculated and observed positive-parity energy levels of $^{16}$C up to
 $J^\pi=12^+$  states. Open boxes show the observed states with the definite spin-parity assignments, and
 other  symbols show the calculated result. The filled circles, triangles  and lines show 
 the ground, triangular  and linear-chain bands, while lines show the states without cluster  structure.}
 \label{fig:spectrum} 
\end{figure*}
Figure \ref{fig:spectrum} shows the spectrum up to $J^\pi=12^+$ state obtained by the GCM calculation
including whole basis wave functions. We classified the obtained states to the 'ground band', 'triangular
band', 'linear-chain band' and other non-cluster states based on their $B(E2)$ strengths and the overlap
with the basis wave functions. Table \ref{tab:band} shows the member
states of these bands with small angular momenta.  
\begin{table}[h]
 \caption{Excitation energies (MeV) and  proton and neutron root-mean-square radii (fm) of several member
 states  of the 'ground band', 'triangular band' and 'linear-chain band'. Numbers in the parenthesis are the
 observed data.}\label{tab:band} 
\begin{center}
 \begin{ruledtabular}
  \begin{tabular}{lcccc} 
   band & $J^\pi$ & $E_x$ & $r_{p}$ & $r_{n}$\\
   \hline
   ground & $0^+_1$ & 0.0 & 2.61 & 2.84 \\
               & $2^+_1$ & 1.3 (1.77) & 2.60 & 2.83 \\
               & $4^+_1$ & 3.9 (4.14) & 2.56 & 2.77 \\
   \hline
   triangular & $0^+_2$ & 8.0 & 2.75 & 3.09 \\
   $K^\pi=0^+$     & $2^+_4$ & 9.4 & 2.74 & 3.08 \\
                   & $4^+_4$ & 12.7 & 2.76 & 3.06 \\
   \hline
   triangular & $2^+_5$ & 10.1 & 2.74 & 3.08 \\
   $K^\pi=2^+$     & $3^+_3$ & 11.7 & 2.74 & 3.07 \\
                   & $4^+_6$ & 13.7 & 2.74 & 3.08 \\
   \hline
   linear-chain & $0^+_5$ & 15.5 & 3.54 & 3.71 \\
                     & $2^+_9$ & 15.9 & 3.14 & 3.27 \\
                     & $2^+_{10}$ & 16.3 & 3.38 & 3.54 \\
                     & $4^+_{11}$ & 17.6 & 3.22 & 3.38 \\
                     & $4^+_{12}$ & 17.8 & 3.21 & 3.39 \\
  \end{tabular}
 \end{ruledtabular}
 \end{center}
\end{table}

The member states of the ground band shown by circles in Fig. \ref{fig:spectrum} are dominantly composed of
the basis wave functions with $(sd)^2$ configuration on the energy curve. The ground state has the largest
overlap with the basis wave function shown in Fig. \ref{fig:density} (a)(b) that amounts to 0.95, and the
calculated binding energy is $-110.6$ MeV that nicely agrees with the observed value ($-110.8$ MeV). The
excitation energies of the $2^+_1$ and $4^+_1$ states are also reasonably 
described. However our result considerably overestimates the observed $B(E2;2^+_1\rightarrow 0^+_1)$
strength reported by experiments \cite{imai04,ong06,ong08,wied08,petr12} that ranges from 0.92 to 4.2
$e^2\rm fm^4$. There have been many discussions about the possible hindrance \cite{yeny05,petr12,saga04} of
$B(E2)$, and in the case of the AMD study \cite{yeny05}, the origin of the hindrance was attributed to the
different proton and neutron deformation. On the other hand, the present results does not describe it as
mentioned before, and it leads to the overestimation of $B(E2)$ (Tab. \ref{table:be2}).  

Owing to its triaxial deformed shape, the triangular configuration generates two rotational
bands built on the $0^+_2$ and $2^+_5$ states. We call them $K^\pi=0^+$ and $2^+$ bands in the following,
although the mixing of the $K$ quantum number in their  GCM wave functions (Eq. (\ref{eq:gcmwf})) is
not negligible. Compared to the linear-chain state, these bands have less pronounced clustering and $\alpha$ 
clusters are considerably distorted, therefore the band head energies are well below the cluster
thresholds. The member states have large overlap with the basis wave function shown in
Fig. \ref{fig:density} (c)(d) which amount to, for example, 0.93 in the case of the $0^+_2$ state. However,
the member states with larger angular momentum with $J^\pi\geq 5^+$ are fragmented  into several states due
to the coupling with other non-cluster configurations. The fragmentation gets stronger as the angular
momentum increases, and hence the member states with $J^\pi\geq 9$ and band terminal are unclear. Due
to larger deformation of the triangular states, the inter- and intra-band $B(E2)$ strengths
between the $K^\pi=0^+$ and  $K^\pi=2^+$ bands are enhanced compared to the ground band.

\begin{table}[h]
\caption{The calculated intra- and inter-band $B(E2;J_i\rightarrow J_f)$ ($e^2\rm fm^4$) strengths for
 low-spin member states of the ground, triangular and linear-chain bands. Transitions less than 5 $e^2\rm
 fm^4$ are not shown.}
\label{table:be2}
\begin{center}
 \begin{ruledtabular}
 \begin{tabular}{lcc}
   & $J_i\rightarrow J_f$ & $B(E2;J_i\rightarrow J_f)$ \\ \hline
  ground $\rightarrow$ ground& $2^{+}_{1}\rightarrow 0^{+}_{1}$ & 6.0 \\
  & $4^{+}_{1} \rightarrow 2^{+}_{1}$ & 5.1 \\ \hline
  triangular & $2^{+}_{4} \rightarrow 0^{+}_{2}$ & 10.9 \\
  $K^\pi=0^+$ $\rightarrow$ $K^\pi=0^+$ & $4^{+}_{4}\rightarrow 2^{+}_{4}$ & 15.7 \\\hline
  triangular & $3^{+}_{3} \rightarrow 2^{+}_{5}$ & 17.9 \\
  $K^\pi=2^+$ $\rightarrow$ $K^\pi=2^+$ & $4^{+}_{6}\rightarrow 3^{+}_{3}$ & 9.5 \\
                                        & $4^{+}_{6}\rightarrow 2^{+}_{5}$ & 6.2 \\\hline
  triangular & $2^{+}_{5} \rightarrow 0^{+}_{2}$ & 6.9 \\
  $K^\pi=2^+$ $\rightarrow$ $K^\pi=0^+$ & $3^{+}_{3}\rightarrow 2^{+}_{4}$ & 10.4 \\
                                        & $3^{+}_{3}\rightarrow 4^{+}_{4}$ & 8.3 \\\hline

  linear-chain $\rightarrow$ linear-chain & $2^{+}_{9}\rightarrow 0^{+}_{5}$ & 58.9\\
                                          & $2^{+}_{10}\rightarrow 0^{+}_{5}$ & 182.4\\
                                          & $4^{+}_{11}\rightarrow 2^{+}_{9}$ & 114.3\\
                                          & $4^{+}_{11}\rightarrow 2^{+}_{10}$ & 70.8\\
                                          & $4^{+}_{12}\rightarrow 2^{+}_{9}$ & 29.5\\
                                          & $4^{+}_{12}\rightarrow 2^{+}_{10}$ & 158.0\\
 \end{tabular}
 \end{ruledtabular}
\end{center}
\end{table}

The  linear-chain configuration generates a rotational band built on the $0^+_5$ state at 15.5 MeV,
that is close to the $^{4}$He+$^{12}$Be and $^{6}$He+$^{10}$Be threshold energies. The band head state
$0^+_5$ has the largest overlap with the basis wave function shown in Fig.\ref{fig:density} (e)(f) which
amounts to 0.92, but the member states with $J^\pi=2^+$, $4^+$ and $6^+$ are fragmented into two states due
to the coupling with other non-cluster basis wave functions. For example, the $2^+_9$ and $2^+_{10}$ states
respectively have 0.30 and 0.65 overlaps with the basis wave function of Fig.\ref{fig:density} (e)(f). By
averaging the excitation energies of the fragmented member states, the moment-of-inertia is estimated as
$\hbar/2\Im=112$ keV. Because of this strong deformation comparable with hyperdeformation, the member states
has huge intra-band $B(E2)$ that is about several tens times as large as those in other bands.  Naturally,
as the angular momentum increases, the excitation energy of the linear-chain state is lowered relative to
other structures, and the $J^\pi=10^+$ member state at $E_x=27.8$ MeV becomes the yrast state. Different
from the triangular band, the high-spin member states with $J^\pi \geq 8^+$ are not fragmented and the band
structure looks rather clear. Since the excitation energy of the high-spin state with linear-chain
configuration is relatively lower than others, the coupling with the non-cluster states and hence the
fragmentation of the 
states may be hindered. Thus, we predict the stable  linear-chain configuration with
molecular-orbits whose band-head energy is around $^{4}$He+$^{12}$Be and $^{6}$He+$^{10}$Be
thresholds. Owing to its large moment-of-inertia, the $J^\pi=10^+$ member state becomes an yrast
state. Those suggest that the linear-chain band might be populated in the $^{4}$He+$^{12}$Be and
$^{6}$He+$^{10}$Be reaction channels.

\section{SUMMARY}
We have studied 3$\alpha$ cluster states of $^{16}$C based on the AMD calculations. By the variational
calculation  with the constraint on the quadrupole deformation parameter $\beta$, it was found that two
different 3$\alpha$ cluster states appear depending on the magnitude of the deformation and the valence
neutron configurations. The  triangular configuration of 3$\alpha$ clusters is accompanied by the valance
neutrons in a   $(sd)^4$ configuration, while the linear-chain configuration has the valence neutrons with a
$(sd)^2(pf)^2$  configuration. From the analysis of the neutron single particle orbits, it is shown that
the valence 
neutron orbits of the linear-chain configuration is understood well in terms of molecular-orbits and it
is qualitatively in good accordance with the $(3/2^-_\pi)^2(1/2^-_\sigma)^2$ configuration suggested by
molecular-orbital model. We also pointed out parity asymmetry of the linear-chain configuration that
originates in 
$^{10}{\rm Be}$+$\alpha$+$2n$ cluster nature. The GCM calculation demonstrated that the wave function of the
linear-chain state is well confined within a region of small $\gamma$, and hence, it is stable against
bending motion. We predict the presence of rotational bands associated with 3$\alpha$ cluster states. In
particular, the linear-chain band is built in the vicinity of the $^{4}$He+$^{12}$Be and $^{6}$He+$^{10}$Be
thresholds energies, and the $J^\pi=10^+$ state becomes an yrast state.

\acknowledgements
Part of the numerical calculations were performed on the HITACHI SR16000 at KEK. One of the authors (M.K.) 
acknowledges the support by the Grants-in-Aid for Scientific Research on Innovative Areas from MEXT
(Grant No. 2404:24105008) and JSPS KAKENHI Grant 563 No. 25400240.

\end{document}